\documentclass[12pt]{article} 
\usepackage{epsfig}
\def\i129{\rm{^{129}I}}

\begin{document}

\hfill AS-TEXONO/05-10 \\
\hspace*{1cm} \hfill \today

\begin{center}
\Large  
{\bf
Measurement of Trace $^{129}$I Concentrations in\\
CsI Powder and Organic Liquid Scintillator\\ 
with Accelerator Mass Spectrometry
}\\[2ex]
\large
K.J.~Dong$^{a,b}$,
M.~He$^{a}$,
S.~Jiang$^{a}$,
H.T.~Wong$^{b,}$\footnote{Corresponding~author $-$\\
\hspace*{1cm} Mailing Address:
Institute of Physics, Academia Sinica, Taipei 11529, Taiwan;\\
\hspace*{1cm} Email:~htwong@phys.sinica.edu.tw;
Tel:+886-2-2789-6789;
FAX:+886-2-2788-9828.},
J.Z.~Qiu$^{a,c}$,\\
Y.J.~Guan$^{a}$,
S.H.~Li$^{a}$,
S.Y.~Wu$^{a}$,
M.~Lin$^{a}$,
Q.B.~You$^{a}$,\\	
Y.W.~Bao$^{a}$,
Y.M.~Hu$^{a}$,
D.~Zhou$^{a}$, 
X.Y.~Yin$^{a}$,
J.~Yuan$^{a}$,

\end{center}
\normalsize

\begin{flushleft}
{$^{a}$\rm 
Department of Nuclear Physics,
Institute of Atomic Energy, Beijing 102413.}\\
%% 102413, China.\\}
{$^{b}$\rm
Institute of Physics, Academia Sinica, Taipei 11529.}\\
%% 115, Taiwan.\\}
{$^{c}$\rm
Armed Police Force Academy, Langfang 065000.}
\end{flushleft}

\vspace*{0.1cm}

\begin{abstract}

Levels of trace radiopurity in active
detector materials is a subject of major concern 
in low-background experiments.
Procedures were devised to measure
trace concentrations of $\i129$ in the inorganic salt
CsI as well as in organic liquid scintillator
with Accelerator Mass Spectrometry (AMS) which leads to
improvement in sensitivities by several orders of
magnitude over other methods.
No evidence of their existence in these materials
were observed. Limits of 
$< 6 \times 10^{-13}$~g/g and 
$< 2.6 \times 10^{-17}$~g/g 
on the contaminations of $\i129$ 
in CsI and liquid scintillator, respectively, 
were derived.
These are the first
results in a research program whose goals
are to develop techniques to measure trace
radioactivity in detector materials by
AMS.

\end{abstract}

\begin{flushleft}
\end{flushleft}

\vfill

%%\begin{center}
%%{\it ( In preparation for Nucl. Instrum. Methods B ) }
%%\end{center}

\pagebreak

\section{Introduction}

\begin{figure}
\centerline{
\epsfig{file=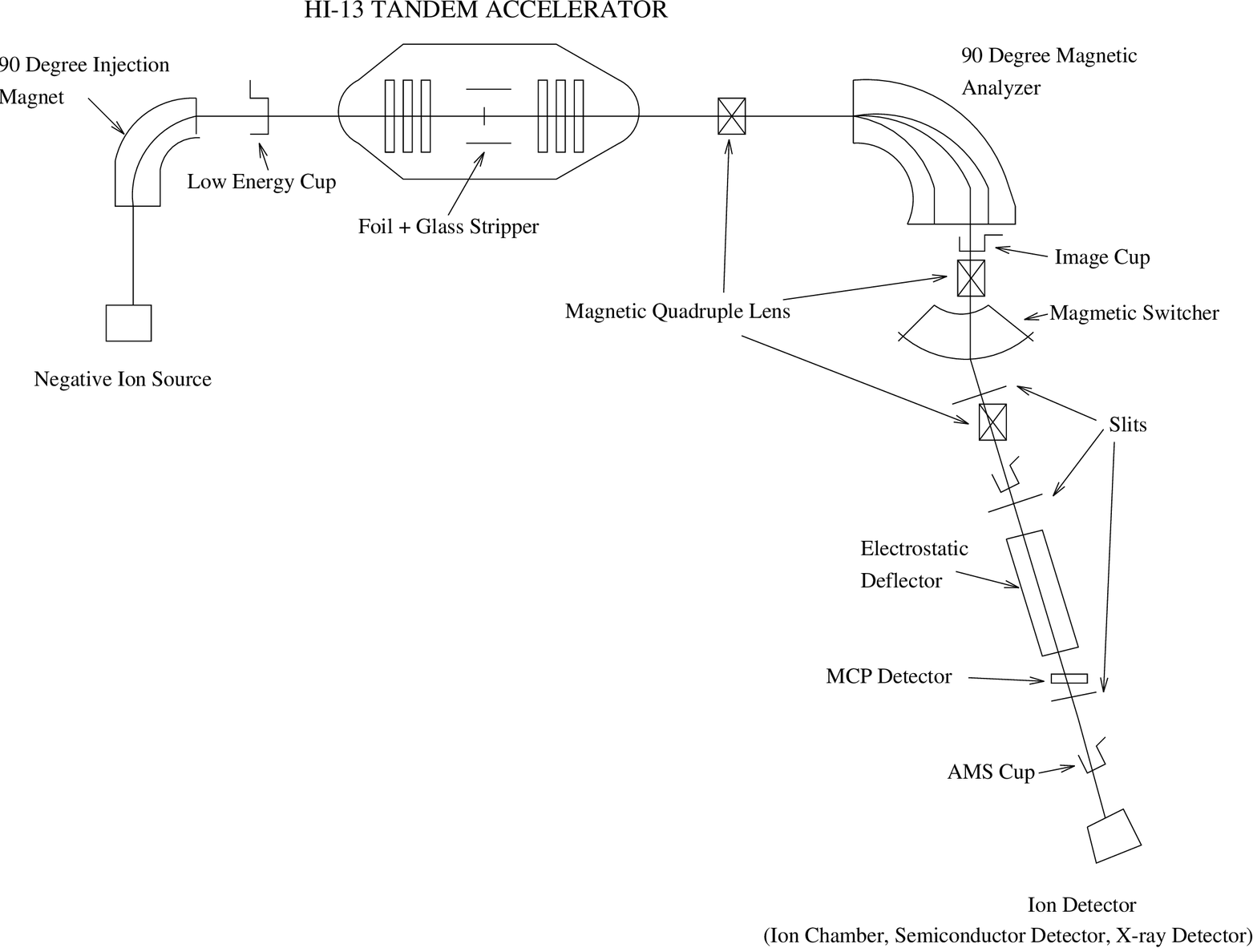,width=13cm,angle=270}
}
\caption{
Schematic layout of the
13~MV Tandem Accelerator Facility at CIAE
for AMS.
}
\label{ciaeams}
\end{figure}

Measurement of trace concentrations of naturally-occurring
and cosmic-ray induced  radioactive isotopes is
an important technique with 
major impact to low background experiments,
such as those for Dark Matter searches, as
well as the
studies of double beta decays, 
reactor and solar neutrinos~\cite{pdg04}.
The TEXONO Collaboration is pursuing 
a research program in low energy 
neutrino and astroparticle physics~\cite{texono}.
One of the efforts is to perform such
trace radiopurity measurements using
the techniques 
of Accelerator Mass Spectrometry (AMS)~\cite{amsoverview,amsradio},
which can potentially lead to significant
improvements in sensitivities and flexibilities
over existing methods.
In this article,
we reported on the measurement of trace 
$\i129$ in inorganic crystal
and organic liquid scintillators
with the AMS facility at the 
China Institute of Atomic Energy (CIAE)~\cite{ciaeamsref},
shown schematically in Figure~\ref{ciaeams}.

The isotope $\i129$ is 
a long-lived (half-life $1.57 \times 10^7$~years)
fission product with yields of
0.74\% and 1.5\% for
thermal neutron-induced fissions of
$^{235}$U and $^{239}$Pu, respectively.
It is commonly
found in the environment, iodine
being readily soluble in water.
Measurements of trace $\i129$ concentration
are adopted world-wide for nuclear safeguards,
in the detection and prevention of accidental or
deliberate discharge of nuclear waste
debris into the environment~\cite{nwaste}.
Another application 
is on radioactive dating of, for instance,
oil field materials~\cite{dating}.

The isotope $\i129$ decays via 
\begin{eqnarray*}
\i129 & ~ \rightarrow ~ & ^{129}Xe^* ~ + ~ e^- ~~ 
( \tau_{\frac{1}{2}} = 1.57 \times 10^7 ~ y ~ ; ~ Q_{\beta}=194~keV )  \\
^{129}Xe^*  & ~ \rightarrow ~ & ^{129}Xe ~ + ~ \gamma  ~~
( \tau_{\frac{1}{2}} = 0.97~ns ~ ; ~ E_{\gamma} = 39.6~keV ) ~~. 
\end{eqnarray*}
Such processes can contribute to the background
in Dark Matter~\cite{pdg04,cdmreview}
and low energy neutrino experiments, such as the
searches of neutrino magnetic moments~\cite{pdg04,numagmom}. 
Concentrations of $\i129$ can also indicate
the contamination levels of other problematic
fission fragments such as $^{137}$Cs 
inherently present in the materials.

The techniques of measuring $\i129$ with AMS 
are by now matured, following early works 
in the 80's~\cite{i129ams}.
They improve over the 
various other measurement methods with
Radio-chemical Neutron Activation 
Analysis~\cite{i129naa},
Inductively Coupled Plasma 
Mass Spectrometry~\cite{i129ms} 
and 
Liquid Scintillation Counting~\cite{i129count}.
The AMS technique has strong rejection
capabilities for isobaric, 
molecular and isotopic interferences,
providing powerful background suppression.
Consequently, AMS is
usually taken as the best among the various
measurement methods,
exceeding the others by three-to-four orders of magnitudes.
It is commonly adopted for the
tasks of environment monitoring.
For instance,
anomalous concentrations of $\i129$ 
in rainwater samples collected shortly after 
the Chernobyl accident were measured by AMS~\cite{chernobyl}.

Measurement of trace radiopurity
in detector materials is a subject
of great importance 
in low-background experiments.
It is usually performed by 
high-purity germanium detectors~\cite{hpge}
or, in the most elaborate case,
with dedicated big-volume liquid
scintillator~\cite{ctf}.
Both of these techniques are not applicable
to $\i129$. We extended the 
list of measurable isotopes to
include $\i129$ by the AMS methods.

An organic liquid and an inorganic
salt were selected for studies since
they require
different experimental procedures
and systematic effect considerations.
The organic liquid studied
is the standard 
mesitylene(1,3,5-trimethylbenzene)+PPO  
liquid scintillator (LS) 
mixture\footnote{Supplier: Gaonengkedi 
Science \& Technology Co. Ltd., China}.
The processing and measurement procedures
with other organic solvent and dyes 
are expected to be very similar.
The inorganic salt selected 
was CsI powder\footnote{Supplier: Chemtall GMBH, Germany},
since CsI(Tl)
crystal scintillators
are being used in 
reactor neutrino~\cite{texonocsi}
and dark matter~\cite{kims} experiments.
Being iodine based, the $\i129$ contaminations
are expected to more likely compared to
the other materials.

\section{Experimental Set-Up Procedures}

In a typical AMS facility,
the samples to be measured
are ionized by a Cs sputtering
negative ion source. The $\i129$
and $^{127}$I ions are
selected and accelerated alternatively.
The $\i129$ ions are eventually
detected by a detector, while
the $^{127}$I current  is
measured by a Faraday cup.

The overall transmission
efficiency common to both isotopes
from the ion source to the detector
%before further signal-over-background enhancement
is about 1\%.  
This was determined with a silver iodide (AgI) sample
by comparing the currents between the ``Low Energy Cup''
and ``AMS Cup'' at the
initial and final stages, respectively,
as depicted in  Figure~\ref{ciaeams}.

We report on the experimental details in the following
sub-sections.

\subsection{Pre-Processing}

No chemical procedures 
are necessary for the CsI powder
which was directly used in the AMS measurement.
However, CsI is a hygroscopic material 
which can easily lead to 
injector magnet excursion. Accordingly,
the CsI samples were deposited quickly 
on to a cathode of electrolytic copper in
a dry box. The cathode was then baked in
an oven at 100$^o$C for two days prior
to the measurement.
 
\begin{figure}
\epsfig{file=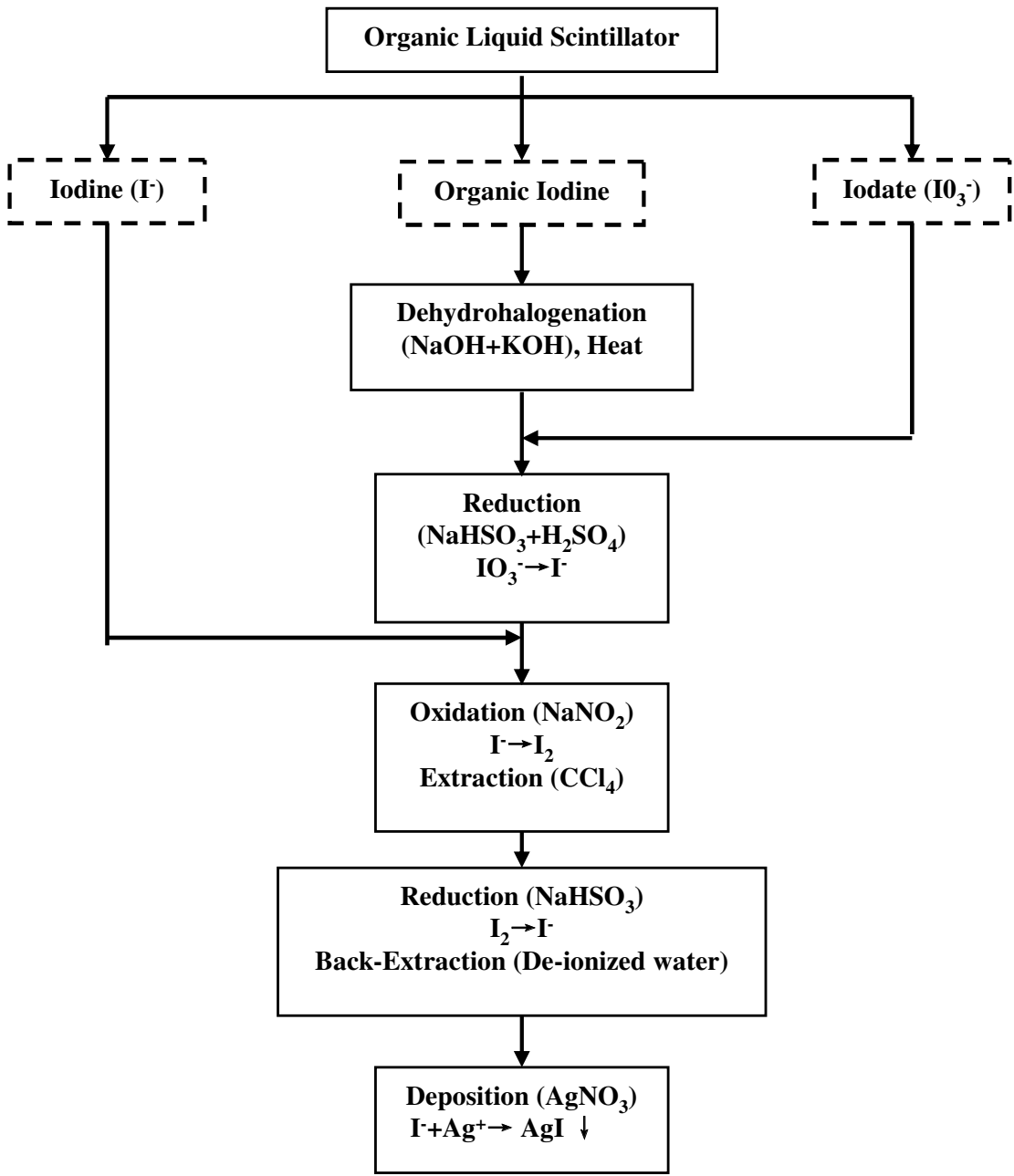,width=15cm}
\caption{
The flow chart of samples preparation for organic liquid scintillator.
}
\label{flowchart}
\end{figure}

For the organic LS, 
an extraction procedure for 
the possible iodine contaminations
has to be devised. 
The adopted 
sample preparation procedure
is shown schematically in 
Figure~\ref{flowchart}.
Similar schemes have been
devised and studied 
in previous work~\cite{iprocess}.
A volume of 100 ml of LS was
evaporated under vacuum. The residuals
left behind consisted mostly of 
the solid PPO powder,
as well as trace concentrations
of the other impurities. 

A KI carrier 
solution of mass 10~mg and 
a solution mixed with  2 mol/l NaOH 
and 2 mol/l KOH in a 3:2 ratio were 
added to the residual solid. 
After stirring to ensure a homogeneous solution,
the mixture was transferred into a crucible and 
put onto a sand bath to dry. 
The dried sample was ashed into a
muffle furnace at 600$^o$C, and then
leached with de-ionized water. 
After being processed by 
a centrifuge, the iodate in the leached
solutions was reduced to iodide with sulfuric acid 
and sodium hydrogen sulfite (NaHSO$_3$). The iodide 
was oxidized to I$_2$ by the addition of sodium 
nitrite and then extracted with 
carbon tetrachloride (CCl$_4$)
and back-extracted 
into de-ionized water by reduction of the 
I$_2$ with 5\% NaHSO3. 

These extraction and back-extraction
steps were repeated until the purple color of 
carbon tetrachloride disappeared. The aqueous 
phase was boiled for a short time to remove 
the residual CCl$_4$. 
After cooling, a silver nitrate (AgNO3) 
solution was added immediately  and
processed with a centrifuge.
The end-product AgI was 
rinsed by de-ionized water, dried and collected. 
Finally, the AgI was mixed with Nb with
a ratio of 1:2 
and kept for subsequent AMS measurement.

As illustrated in Figure~\ref{flowchart},
the three possible forms where iodine may
exist in LS can all be extracted into
the AgI samples for measurement.
The extraction efficiency of 
this procedure was 80\%,
determined by comparing
the ratio, after proper normalizations,
of the mass of the extracted AgI 
compared to that of KI initially introduced.

\subsection{Injection and Accelerator}

The $\i129$  concentration in 
the CsI powder and the LS extracted as AgI
were measured with CIAE-AMS facility
depicted schematically in Figure~\ref{ciaeams}.
The tandem accelerator was operated at a terminal 
voltage of 8.0 MV.  
A ``Multi-Cathode Source of Negative Ions 
by Cesium Sputtering''
%%(MC-SNICS)
was used as the negative ion source. 
%%manufactured by NEC 
Forty samples 
were positioned on the target wheel at one time. 
The wheel could be rotated 
without affecting the vacuum  conditions
such that stable operating configurations were maintained
during measurements of a group of samples. 

The I$^-$ negative ions 
extracted from cesium sputter source were focused 
by a trim einzel lens and a double 
focusing 90$^o$ deflecting magnet where
momentum analysis selected the negative 
ion beams of interest.
The ions were guided to 
an aperture of 2~mm diameter 
located at the entrance of the 
pre-acceleration tube, and
then accelerated up to
about 120 keV  kinetic energy
with a terminal voltage of 
about 8~MV. 
A carbon foil was attached 
at the head of accelerator. The molecular 
background was eliminated due to break-up of 
molecular ions. 

After passing through the accelerator
tank, ions with
charged state 11$^+$ were selected by a 90$^o$ 
double focusing analyzing magnet with a
mass energy product (M$\cdot$E/Z$^2$) at 200 
to suppress the isotopic background. 
A high-resolution electrostatic deflector 
was placed at a branch beam line to 
further reduce the isotopic 
background and other undesired beams.

\subsection{Detector}

\begin{figure}
\epsfig{file=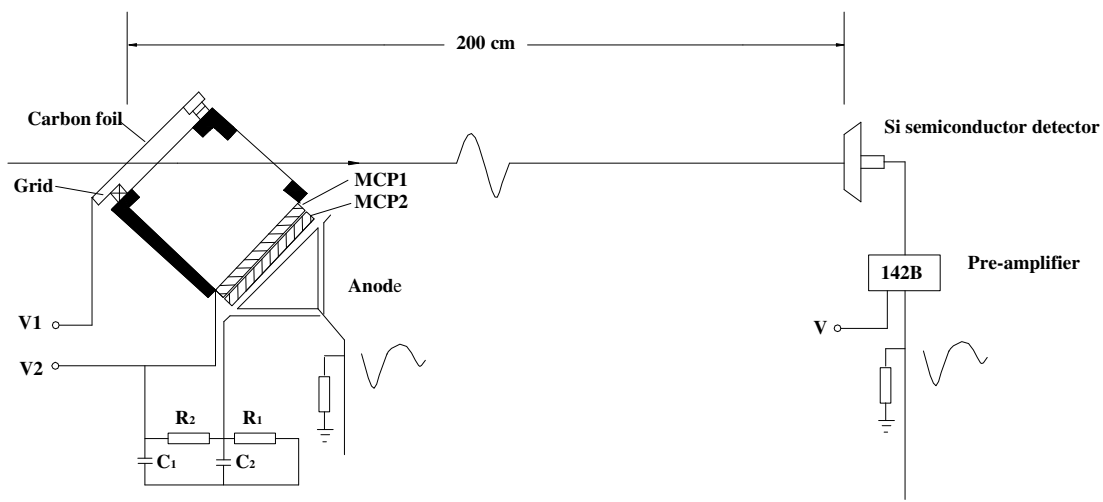,width=15cm}
\caption{
The schematic layout of the Time-of-Flight system.
}
\label{tof}
\end{figure}

Particle detection and identification of 
$\i129$ was
performed via Time-of-Flight (TOF) detector. 
A detailed layout of the TOF system is shown in 
Figure~\ref{tof}. 
A Micro-Channel Plate (MCP)
detector provided the ``$START$'' signal 
while  a
gold-silicon surface barrier detector 
located 200 cm downstream  
was used to to give a ``$STOP$''signal.
The resolution of the TOF system is 600 ps. 
The difference between the flight time of 
$\i129$ and $^{127}$I was
2~ns under the conditions  of equal
momentum at a kinetic energy of about 96~MeV.

\subsection{Calibration and Cross-Checks}

Standard AgI samples with $\i129$ were prepared 
and verified by the 
procedures described in Ref. ~\cite{i129calibrate}. 
%% $\i129$-standard of IAEA-375 reference material
According to the contrast results, an
uncertainty better than 1\% for the 
$\i129$ source strength in the samples
could be measured. 
The samples were subsequently measured
at the AMS facility under the conditions 
discussed. The derived $\i129$/$^{127}$I ratios
in several measurements
were consistent to 10\% with 
the reference values varying from 
$10^{-10}$ to $10^{-12}$.

\begin{figure}
\epsfig{file=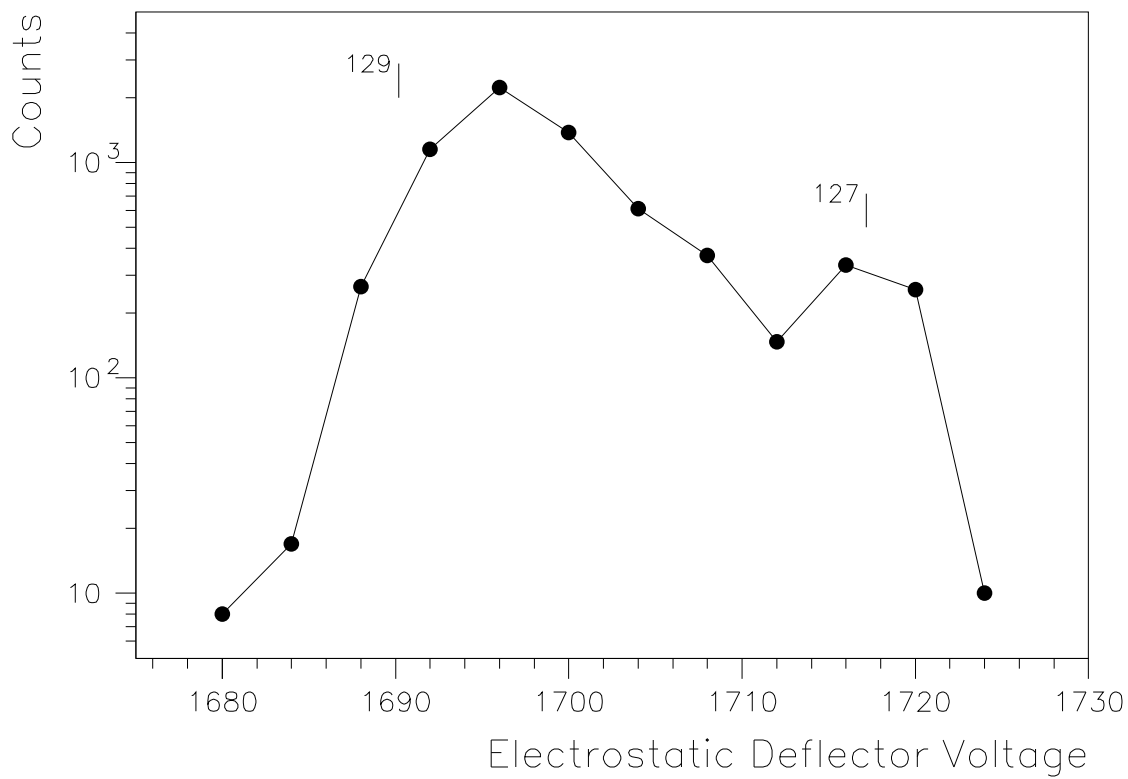,width=15cm}
\caption{
The scan spectra of electrostatic deflector
}
\label{scan}
\end{figure}

\begin{figure}
\epsfig{file=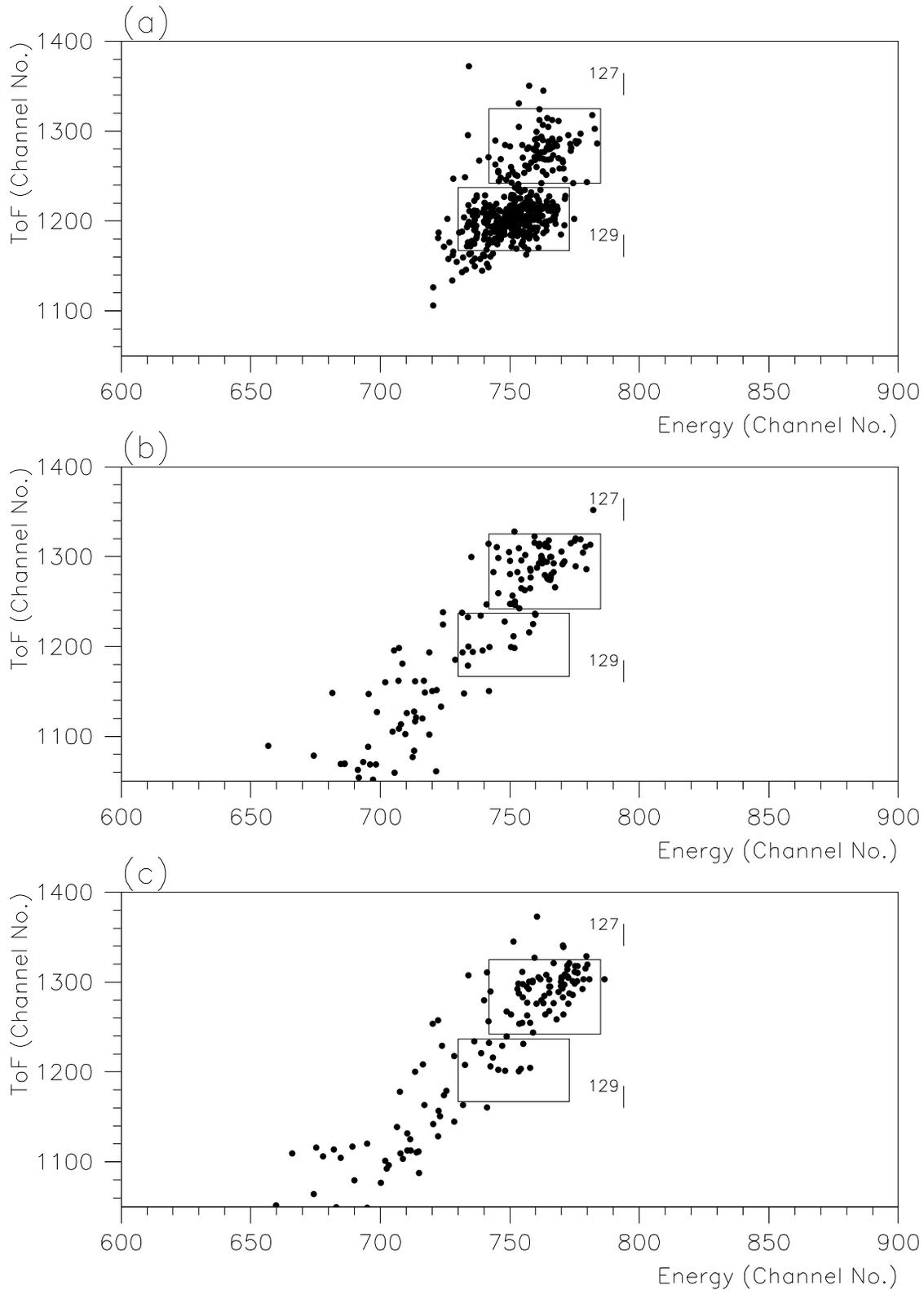,width=15cm}
\caption{
TOF spectra measured by AMS
on a) reference sample at
$\i129$/$^{127}$I=$1.0 \times 10^{10}$,
with which the $\i129$-signal-box can be
defined,
b) CsI powder and 
c) liquid scintillator.
There are no evidence of
excess of $\i129$ above 
the background from
$^{127}$I 
in (b) and (c).
}
\label{tofdata}
\end{figure}

The energy of the $^{127}$I ions are
1.5\% higher than that of $\i129$ ions
at the same momentum.
Accordingly, the $^{127}$I ions
would deflect more than those
of $\i129$ by the electrostatic analyzer,
as depicted in Figure~\ref{scan}, such
that an energy resolution ($\Delta$E/E)
better than 0.5\% was achieved.
The TOF selection suppressed the $^{127}$I
by another two orders of magnitude, as
shown in Figure~\ref{tofdata}a for the
standard sample with known concentration
of $\i129$/$^{127}$I at $1.0 \times 10^{-10}$.
An ``$\i129$-signal-box'' region can be
defined from this measurement to locate
the $\i129$ candidate events.

\subsection{Results}

The suppression factors for $^{127}$I
due to the various AMS components
are summarized in Table~\ref{efftable}. 
An efficient transmission of $\i129$ 
at $\sim$60\%
was achieved, as demonstrated by the
measurements with the AgI
calibration samples with known
absolute strengths and
$\i129$/$^{127}$I ratios.

\begin{table}
\begin{center}
\begin{tabular}{|lcc|}
\hline
Components & $^{127}$I Rejection & $^{129}$I Transmission \\ \hline
Deflecting Magnet & $\sim$$10^{-3}$ & $\sim$1 \\
Analyzing Magnet & $\sim$$10^{-6}$ & $\sim$1 \\
Electrostatic Deflector & $\sim$$10^{-2}$ & $\sim$0.8 \\
TOF Detector & $\sim$$10^{-2}$ & $\sim$0.8 \\ \hline
Total & $\sim$$10^{-13}$ & $\sim$0.6 \\ \hline
\end{tabular}
\end{center}
\caption{
Rejection power of $^{127}$I and transmission efficiency
of $^{129}$I for the various components in
the AMS facility. 
In addition to these factors, 
there is an $\sim$1\% overall transmission efficiency 
applicable to both isotopes.  }
\label{efftable}
\end{table}

The TOF scattered plots for the CsI
and LS samples
are presented in
Figures~\ref{tofdata}b and c, respectively.
In both of these cases, as well as in other
control measurements with commercially available
KI and AgI powder, the measured 
$\i129$-signal-box/$^{127}$I 
ratios are all
$\sim 3 \times 10^{-13}$. 
In comparison, a 
``blank measurement'' of only
the copper cathode {\it without} samples
gave zero counts in the $\i129$-signal-box,
indicating that the events are iodine-related.
These events can be 
due to actual $\i129$ contaminations in the
samples as well as background from spurious
effects or tail distributions of the
dominant $^{127}$I. 
Therefore, conservative limits of 
\begin{displaymath}
\rm{
\i129 / ^{127} I  ~ < ~ 3 \times 10^{-13} 
}
\end{displaymath}
can be derived
in all four cases (CsI, LS with KI as carrier, 
as well as the KI and AgI control samples). 

This limit 
is directly applicable to
characterize the $\i129$ contaminations in CsI. 
It can be alternatively expressed as 
$< 6 \times 10^{-13}$~g/g. 
Accordingly, the $\beta$-decay background
due to $\i129$ in CsI is
less than 83~kg$^{-1}$day$^{-1}$.
In comparison, a recent
measurement of $^{137}$Cs contaminations in CsI(Tl)
crystals~\cite{csibkg} was
$( 1.55 \pm 0.05 ) \times 10^{-17}$~g/g.
Both $^{137}$Cs and $\i129$ are fission fragments
found in the environment, such that their
contamination levels in CsI are expected to be similar.
Measurement sensitivity for $^{137}$Cs is much
enhanced due to its much shorter half-life and
the emissions of mono-energetic 
$\gamma$-rays which are easily identified.

For the LS measurement, the volume of the LS
and the mass of the KI carrier were known,
from which the limit of $\i129$ concentration
in LS of $<2.6 \times 10^{-17}$~g/g can be
derived, implying a background $\beta$-decay
rate in LS
of less than 15~ton$^{-1}$day$^{-1}$.

Further improvement on the sensitivities in
the LS measurements is possible, through the use
of larger initial LS samples as well
as reduced KI mass in the carrier solution.
In addition, 
if the residual events in the 
$\i129$-signal-box 
can be identified to be
background from $^{127}$I  through
more detailed studies of the TOF system response,
the limits can also be improved.

\section{Conclusion}

Measurements on 
the $\i129$ concentrations in an inorganic salt
and organic liquid scintillator were performed
with the AMS techniques.
No evidence were observed for $\i129$ contaminations
and sensitive limits were derived. The limits
are relevant to the design and interpretation
of various low background experiments.

The measurements of $\i129$ are the first 
``demonstration-of-principle'' efforts 
of devising techniques and procedures in
the trace radiopurity measurements 
of naturally-occurring isotopes using AMS. 
Research program on the applications of 
AMS techniques to $^{40}$K and $^{87}$Rb 
are being pursued, while those for
heavier isotopes like $^{238}$U and $^{232}$Th
series are being planned.

The authors gratefully acknowledge 
Y.D~Chen, B.F.~Ni, H.Q.~Tang, W.Z.~Tian
and Z.Y.~Zhou for helpful discussions,
as well as to the technical staff at CIAE 
for smooth accelerator operation.
This work is supported by 
contract 10375099 from the National Natural
Science Foundation, China.  

\pagebreak

\newpage

\end{document}